\documentclass[]{jkas} 

\def\year{2024} 
\def\volume{} 
\def\issue{} 
\def\beginpage{1} 
\def\received{February 10, 2025} 
\def\accepted{March 27, 2025} 
\def\published{---} 
\date{Received \received; Accepted \accepted; Published \published}

\usepackage{flushend} 
\usepackage{multirow}
\usepackage{color}
\usepackage[varg]{txfonts}
\newcommand\ion[2]{{#1}\,{\sc #2}} 

\title{%
Near-Infrared Spectroscopy with IGRINS-2 for Studying Multiple Stellar Populations in Globular Clusters
}

\author[1,$\star$]{Dongwook Lim}{0000-0001-7277-7175}
\author[1]{Young-Wook Lee}{0000-0002-2210-1238}
\author[2]{Sol Yun}{0009-0004-1035-3309}
\author[3]{Young Sun Lee}{0000-0001-5297-4518}
\author[4]{Sang-Hyun Chun}{0000-0002-6154-7558}
\author[4]{Heeyoung Oh}{0000-0002-0418-5335}\author[4]{Jae-Joon Lee}{0000-0003-0894-7824}\author[4]{Chan Park}{0000-0001-9773-3080}\author[4]{Sanghyuk Kim}{}\author[4,5]{Ueejeong Jeong}{}\author[4]{Hye-In Lee}{0000-0003-0871-3665}\author[4]{Woojin Park}{0000-0001-8012-5871}\author[4]{Youngsam Yu}{}\author[4]{Yunjong Kim}{}\author[4]{Moo-Young Chun}{}\author[4]{Jae Sok Oh}{}\author[4]{Sungho Lee}{}\author[4]{Jeong-Gyun Jang}{}\author[4]{Bi-Ho Jang}{}\author[4]{Hyeon Cheol Seong}{}\author[4]{Hyun-Jeong Kim}{0000-0001-9263-3275}\author[5]{Cynthia B. Brooks}{}\author[5]{Gregory N. Mace}{0000-0001-7875-6391}\author[5]{Hanshin Lee}{}\author[5]{John M. Good}{}\author[5]{Daniel T. Jaffe}{0000-0003-3577-3540}\author[4]{Kang-Min Kim}{}\author[4]{In-Soo Yuk}{}\author[4]{Narae Hwang}{0000-0002-2013-1273}\author[4]{Byeong-Gon Park}{0000-0002-6982-7722}
\author[6]{Hwihyun Kim}{0000-0003-4770-688X}\author[7]{Brian Chinn}{}\author[7]{Francisco Ramos}{}\author[7]{Pablo Prado}{}\author[7]{Ruben Diaz}{}\author[8]{John White}{}\author[8]{Eduardo Tapia}{}\author[7]{Andres Olivares}{}\author[7]{Valentina Oyarzun}{}\author[8]{Emma Kurz}{}\author[8]{Hawi Stecher}{}\author[7]{Carlos Quiroz}{}\author[7]{Ignacio Arriagada}{}\author[7]{Thomas L. Hayward}{}\author[8]{Hyewon Suh}{0000-0002-2536-1633}\author[8]{Jen Miller}{}\author[8]{Siyi Xu}{}\author[8]{Emanuele Paolo Farina}{0000-0002-6822-2254}\author[8]{Charlie Figura}{}\author[8]{Teo Mocnik}{0000-0003-4603-556X}\author[8,9]{Zachary Hartman}{0000-0003-4236-6927}\author[8]{Mark Rawlings}{}\author[8]{Andrew Stephens}{}\author[7]{Bryan Miller}{0000-0002-5665-376X}\author[8]{Kathleen Labrie}{}\author[8]{Paul Hirst}{}

\affil[1]{Center for Galaxy Evolution Research \& Department of Astronomy, Yonsei University, 50 Yonsei-ro, Seoul 03722, Republic of Korea}
\affil[2]{Department of Earth Environmental \& Space Science, Chungnam National University, 99 Daehak-ro, Daejeon 34134, Republic of Korea}
\affil[3]{Department of Astronomy and Space Science, Chungnam National University, 99 Daehak-ro, Daejeon 34134, Republic of Korea}
\affil[4]{Korea Astronomy and Space Science Institute, 776 Daedeokdae-ro, Yuseong-gu, Daejeon 34055, Republic of Korea}
\affil[5]{University of Texas at Austin, 2515 Speedway, Stop C1400, Austin, Texas 78712-1205, USA}
\affil[6]{International Gemini Observatory/NSF NOIRLab, 950 N. Cherry Ave., Tucson, AZ 85719, USA}
\affil[7]{International Gemini Observatory/NSF NOIRLab, Casilla 603, La Serena, Chile}
\affil[8]{International Gemini Observatory/NSF NOIRLab, 670 N. A’ohoku Place, Hilo, Hawai’i, 96720, USA}
\affil[9]{NASA Ames Research Center, Moffett Field, CA 94035, USA}

\def\corrauthor{%
D. Lim, \email{dwlim@yonsei.a.kr}
}

\def\runningauthor{%
Lim et al.
}

\def\runningtitle{%
Near-Infrared Spectroscopy for Multiple Stellar Populations in Globular Clusters
}

\def\keywords{%
techniques: spectroscopic ---
stars: abundances ---
globular clusters: general ---
globular clusters: individual: M5 ---
infrared: stars
}

\def\abstracttext{%
Recent advancements in near-infrared (NIR) spectroscopy have opened new opportunities for studying multiple stellar populations in globular clusters (GCs), particularly for newly discovered clusters in the inner Milky Way. 
While optical spectroscopy has traditionally played a primary role in detailed chemical abundance studies of GCs, the increasing discovery of GCs in highly reddened environments underscores the need for robust NIR spectroscopic methods.
To evaluate the utility of high-resolution NIR spectroscopy for studying multiple stellar populations, we observed six stars in M5, a well-studied halo GC, using the recently commissioned IGRINS-2 spectrograph on the Gemini-North telescope. 
Our chemical abundance measurements in the NIR wavelength range show good agreement with those derived from high-resolution optical spectroscopy, with minor systematic offsets in elements such as Na and Mg. 
In addition, the measured chemical abundance ratios clearly reproduce the distinctive patterns of multiple stellar populations, including the Na-O anti-correlation.
The ability of NIR spectroscopy to measure C, N, and O abundances with high precision further enhances its utility for studying chemical properties of stars and GCs. 
Our findings demonstrate that IGRINS-2 and similar instruments have significant potential to advance our understanding of GC formation, stellar chemical evolution, and the evolutionary history of the Milky Way.
}

\begin{document}
\jkashead 


\section{Introduction} \label{sec:intro}
The phenomenon of multiple stellar populations in Milky Way globular clusters (GCs) has been a subject of extensive investigation over the past two and a half decades \citep[see][and references therein]{Bastian2018, Gratton2019, Milone2022}. 
Significant progress has been made through observational, theoretical, and simulation-based approaches to identify the unique characteristics of these populations and to understand their origin.
While various signatures of multiple stellar populations can be detected through photometric observations, such as splits in the red giant branch (RGB) and the main sequence \citep[e.g.,][]{Milone2012, Lim2015}, spectroscopic observations are essential for revealing their detailed chemical abundance patterns.

The most distinctive chemical signature of multiple stellar populations, as revealed by spectroscopic observations, is the Na-O anti-correlation. 
This feature has been widely employed to investigate the origins of GCs and their stellar populations, both before and after its establishment by \citet{Carretta2009a, Carretta2009b} as a general characteristic of GCs with multiple stellar populations. 
In addition to the Na-O anti-correlation, other chemical anomalies, such as the Mg-Al and N-C (or CN-CH) anti-correlations, have been identified as common and significant features of GCs \citep[e.g.,][]{Kayser2008, Lim2017, Mucciarelli2018}. 
Typically, later-generation stars are enriched in Na, Al, and N, while being depleted in O, Mg, and C compared to earlier-generation stars within the same cluster.

The prevailing explanation for the chemical differences between multiple stellar populations is that later-generation stars formed from gas polluted by the ejecta of earlier-generation stars. 
Several types of polluting stars have been proposed as potential sources, including asymptotic giant branch stars \citep{D'Antona2004}, fast-rotating massive stars \citep{Decressin2007}, massive interacting binaries \citep{deMink2009}, and stellar winds from massive stars \citep{Kim2018}. 
Although the exact origin of multiple stellar populations in GCs remains unresolved, their distinctive chemical properties provide valuable insights into the contribution of GC-originated stars to the formation and evolution of the Milky Way \citep[see, e.g.,][]{Koch2019, Lim2021, Hong2024}. 
These studies highlight that studying multiple stellar populations in GCs is essential not only for understanding their origins but also for advancing our knowledge of the Galactic formation and evolution.

One of the most active areas of research on Milky Way GCs is the discovery of new stellar clusters in the inner Galaxy, particularly within the bulge region. 
Recent photometric surveys have identified over one hundred cluster candidates, many of which have been confirmed as GCs \citep[][and references therein]{Bica2024, Garro2024}. 
These discoveries have been enabled by the capabilities of near-infrared (NIR) observations, which are less affected by interstellar reddening. 
While NIR photometric surveys have successfully identified new GCs \citep[e.g.,][]{Gran2022}, detailed chemical characterization of these clusters requires high-resolution NIR spectroscopic observations \citep[e.g.,][]{Kunder2020, Fernandez-Trincado2022}.

Many of these newly discovered GCs are located in the inner regions of the Galaxy, making their properties and stellar populations critical for understanding the formation and evolution of the Milky Way bulge. 
However, our current understanding of the use of NIR spectroscopy to study multiple stellar populations in GCs remains limited due to the relative scarcity of high-resolution NIR spectroscopic studies compared to optical ones. 
To date, most chemical properties of stars in GCs have been examined through high-resolution spectroscopy in the optical wavelength range.
Although the Apache Point Observatory Galactic Evolution Experiment \citep[APOGEE;][]{Majewski2017} has provided NIR spectroscopic data for stars in some GCs, its observations are restricted to regions outside the innermost parts of GCs and are limited to the H-band (1.5 $-$ 1.7 $\mu$m). 
This underscores the need to validate the strengths and feasibility of high-resolution NIR spectroscopy for studying multiple stellar populations, particularly through individual observations that cover a broader wavelength range. 
Several instruments with such capabilities already exist, including GIANO \citep{Origlia2014, Caffau2019}, WINERED \citep{Ikeda2016}, CRIRES+ \citep{Follert2014}, and IGRINS \citep{Park2014}. 
These instruments have increasingly been used to study a diverse range of Milky Way stars.
Comparing NIR observations with those obtained from optical spectroscopy will provide valuable guidelines for studying GCs located in regions of severe extinction and will further advance our understanding of GC stellar populations and their role in Galactic evolution.

In this study, we utilize high-resolution NIR spectroscopic data obtained with the Immersion GRating INfrared Spectrograph 2 (IGRINS-2) to evaluate the reliability of NIR spectroscopy for investigating multiple stellar populations in GCs. 
To achieve this, we observed M5 (NGC~5904), one of the most extensively studied GCs across various observational techniques.
\citet{Smith1997} identified a CN-CH anti-correlation among several giant stars in M5, while \citet{Ivans2001} reported significant star-to-star variations in O, Na, and Al abundances.
\citet{Koch2010} further confirmed the presence of multiple stellar populations through distinct Na-O abundance pattern.
More recently, \citet{Lee2017, Lee2021} employed narrow-band photometry to reveal distinct stellar populations characterized by differences in CN and CH strengths.
In addition, \citet{Carretta2009a, Carretta2009b} conducted a systematic study of Na-O and Mg-Al anti-correlations in numerous GCs, including M5, establishing these features as fundamental signatures of multiple stellar populations.
Since M5 has been analyzed in terms of its detailed chemical abundances from high-resolution optical spectroscopy, it serves as an ideal testbed for evaluating the effectiveness of NIR spectroscopy in characterizing multiple stellar populations in GCs.

The structure of this paper is as follows.
The observations and data reduction procedures are described in Section~\ref{sec:obs}. 
Section~\ref{sec:spec} outlines the spectroscopic analysis and chemical abundance measurements. 
The results are presented in Section~\ref{sec:result}. 
Finally, the advantages and limitations of NIR spectroscopy are discussed in Section~\ref{sec:discussion}.

\begin{table*}
\caption{Target information}
\label{tab:target} 
\centering                                    
\begin{tabular}{cccccccc}  
\hline\hline 
\multirow{2}{*}{ID} & R.A.          & Decl.         & $K$       & Exposures     & SNR$_{H}$         & SNR$_{K}$         \\ 
                    & [hh mm ss]    & [dd mm ss]    & [mag]     & [second]      & [pixel$^{-1}$]    & [pixel$^{-1}$]    \\
\hline
900073              & 15 18 32.630  & +02 02 20.45  & 10.204    & 50 $\times$ 4 &  89               &  75               \\ 
900078              & 15 18 04.093  & +02 05 55.86  & 10.225    & 50 $\times$ 4 &  63               &  57               \\ 
900085              & 15 18 44.500  & +02 02 05.77  & 10.304    & 50 $\times$ 4 &  65               &  58               \\ 
900106              & 15 18 18.997  & +02 08 19.14  & 10.548    & 60 $\times$ 4 &  80               &  78               \\ 
900152              & 15 18 12.636  & +02 03 10.58  & 10.938    & 90 $\times$ 4 &  91               &  78               \\ 
900216              & 15 18 24.304  & +02 01 09.53  & 11.453    &150 $\times$ 4 &  79               &  56               \\ 
\hline
\end{tabular}
\end{table*}

\section{Observation and Data Reduction} \label{sec:obs}
\subsection{IGRINS-2} \label{sec:sub:igrins}
IGRINS-2 is a high-resolution NIR spectrograph recently commissioned on the Gemini-North telescope \citep{Lee2022, Oh2024}. 
It is the successor to the original IGRINS \citep{Park2014, Mace2018}, which was previously installed as a visitor instrument at the Gemini-South telescope and the Lowell Discovery Telescope, and is now in operation at the Harlan J. Smith Telescope at McDonald Observatory. 
Both IGRINS and IGRINS-2 provide nearly identical spectral capabilities, offering a spectral resolving power of $R \sim 45,000$ with a slit size of $0.33^{\prime\prime} \times 5^{\prime\prime}$.
IGRINS-2 is equipped with two Teledyne HAWAII-2RG spectrograph detector arrays, enabling simultaneous coverage of the H-band (1.40 $-$ 1.80 $\mu$m) and K-band (1.96 $-$ 2.46 $\mu$m) wavelength ranges.
The instrument began its commissioning phase at Gemini North in late 2023 and is available in shared-risk mode for the 2025A semester at Gemini Observatory.

\subsection{Observation} \label{sec:sub:obs}
The observations were conducted during the IGRINS-2 commissioning test run in April 2024 (Program ID: GN-2024A-ENG-142).
To validate the reliability of NIR spectroscopy for studying multiple stellar populations in GCs, we selected six target stars in M5, a GC that has been well studied using optical spectroscopy. 
\citet{Carretta2009a} measured the chemical abundances of Fe, O, Na, Mg, Al, and Si using high-resolution UVES spectra obtained with FLAMES on the VLT-UT2 telescope.

Among the six targets, three stars (900073, 900078, and 900152) belong to the chemically enriched, later-generation, characterized by enhanced Na and depleted O abundances. 
The remaining three stars (900085, 900106, and 900216) represent the primordial, earlier-generation. 
These stars exhibit clear Na-O anti-correlation based on prior optical spectroscopic studies. 
Furthermore, their bright K-band magnitudes (10.0 $-$ 11.5 mag) made them ideal candidates for the IGRINS-2 commissioning test.

Observations for each target followed an ABBA nod sequence. 
Data for stars 900073, 900106, and 900152 were collected on April 28, 2024, while observations for 900078, 900085, and 900216 were conducted on April 29, 2024. 
The observing conditions on April 28 were Image Quality (IQ) = 85$\%$ and Cloud Cover (CC) = 70$\%$, while those on April 29 were IQ = 70$\%$ and CC = 80$\%$.
Blank sky exposures were taken before and after the target observations, and an A0V-type standard star was observed alongside the targets. 
Total exposure times for each star ranged from 200 to 600 seconds, depending on the brightness of target.
Table~\ref{tab:target} lists the target IDs, coordinates, magnitudes, and exposure details. 
The target IDs used in this study are consistent with those in \citet{Carretta2009a}.

\subsection{Data Reduction} \label{sec:sub:reduction}
\begin{figure*}[t!]
\centering
   \includegraphics[width=0.99\textwidth]{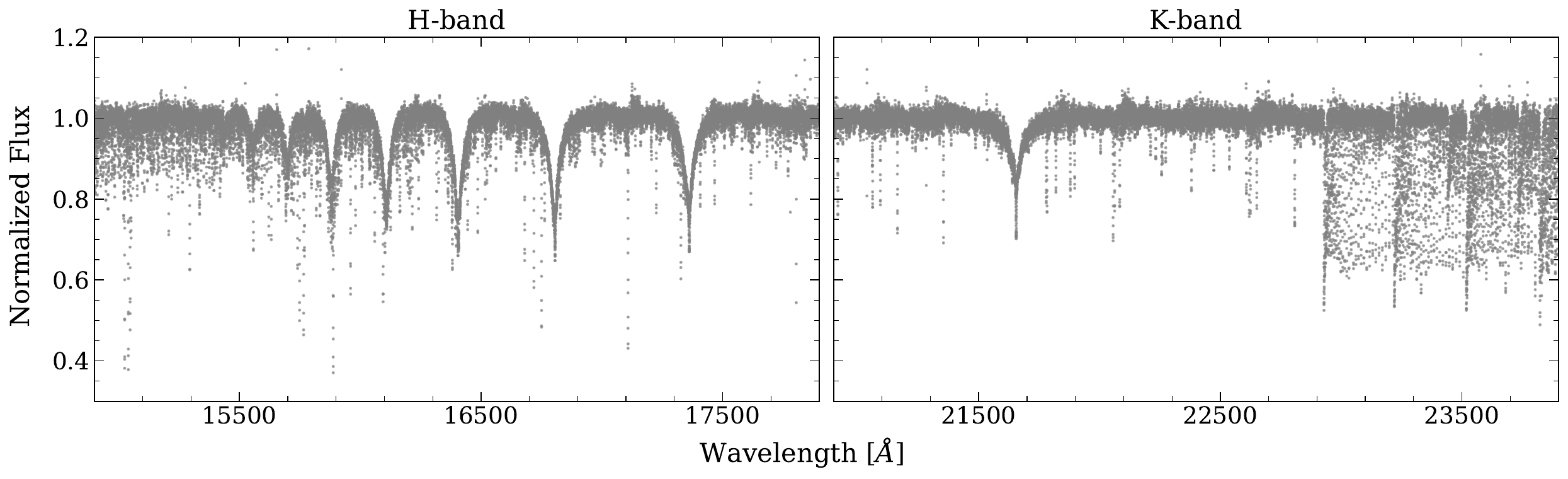}
     \caption{
     Example spectrum of IGRINS-2 in the H- and K-bands for the target star 900073. 
     }
     \label{fig:sed}
\end{figure*}

The data reduction was carried out using the {\em igrins2-dev} branch of the IGRINS Pipeline Package Version 3.0 \citep[PLP;][]{Kaplan2024}, which is specifically optimized for IGRINS-2 data. 
The PLP includes a series of processing steps such as flat-fielding, bad-pixel correction, flexure correction, telluric correction, sky background subtraction, and wavelength solution determination, ultimately producing one-dimensional spectra for each diffraction order. 
These spectra were subsequently combined into a continuous spectrum and continuum-normalized using the IRAF {\em continuum} task.
The data reduction procedure closely follows that used for IGRINS datasets as described in \citet{Lim2022} and \citet{Lim2024a}.
Starting from PLP Version 3.0, a flexure correction procedure has been implemented to enhance the accuracy of the wavelength solution.
Since the wavelength solution of IGRINS-2 is derived from emission lines—primarily OH lines—obtained from a separate sky frame, it can be affected by instrumental flexure.
The newly incorporated flexure correction process mitigates this effect, ensuring a more precise wavelength solution.
An example of the reduced spectrum for 900073 is shown in Figure~\ref{fig:sed}.

The signal-to-noise ratio (SNR) for each target was measured separately for the H- and K-band regions. 
SNRs were determined near 1.63~$\mu$m for the H-band and 2.20~$\mu$m for the K-band. 
As listed in Table~\ref{tab:target}, the SNR is generally higher in the H-band compared to the K-band. 
It should be noted that the achieved SNRs are lower than the estimates provided by the IGRINS-2 integration time calculator, as these observations were conducted during the commissioning test phase.

\section{Spectroscopic Analysis} \label{sec:spec}
\subsection{Radial Velocity} \label{sec:sub:rv}
Radial velocities (RVs) for each star were measured using the cross-correlation method, with synthetic NIR spectra from the Pollux database \citep{Palacios2010} serving as a reference. 
The IRAF {\em fxcor} and {\em rvcor} tasks were utilized to estimate RVs and apply heliocentric corrections. 
RVs were measured separately for the full spectral ranges of the H- (1.40 $-$ 1.80 $\mu$m) and K-bands (1.96 $-$ 2.46 $\mu$m), and their mean value was adopted as the final RV estimate.
The RV uncertainties, provided by {\em fxcor}, account for both the cross-correlation peak fitting error—determined from the center parameter of the fitting function—and the formal uncertainty arising from the peak width and height, following the methodology of \citet{Tonry1979}.
A comparison of our RV measurements with those from \citet{Carretta2009a} and Gaia DR3 \citep{GaiaCollaboration2023} is presented in Table~\ref{tab:rv}.

The RV values derived from the three sources are generally consistent, with differences within a standard deviation of 1.5~km~s$^{-1}$. 
While our RV estimates are slightly higher on average than those reported by \citet{Carretta2009a}, they show closer agreement with the Gaia DR3 values. 
These discrepancies are considered negligible, as they are smaller than the measurement uncertainties and the standard deviation of RVs among cluster member stars.

We note that our RV estimates may exhibit a minor systematic discrepancy (less than 0.5~km~s$^{-1}$) due to temperature stability issues with the immersion grating during the commissioning run.
However, we expect that this issue has been largely mitigated by the flexure correction procedure (see Section~\ref{sec:sub:reduction}).
During flexure correction, each frame undergoes a shift of approximately 0.05 to 0.35 pixels, resulting in an RV difference of 0.15~km~s$^{-1}$ to 0.41~km~s$^{-1}$.
Consequently, more precise RV measurements can be achieved with IGRINS-2 by leveraging improved wavelength calibration using synthetic telluric templates, which account for atmospheric absorption lines \citep[see][]{Stahl2021}, along with flexure correction.

\begin{table}
\caption{Comparison of heliocentric RV measurements}
\label{tab:rv} 
\centering                                    
\begin{tabular}{cccc}  
\hline\hline 
\multirow{3}{*}{ID} & \multicolumn{3}{c}{RV}                                    \\
\cline{2-4}
                    & This study            & Carretta+2009a    & Gaia DR3      \\
                    & [km~s$^{-1}$]         & [km~s$^{-1}$]     & [km~s$^{-1}$] \\
\hline
900073              & 56.60 $\pm$ 0.58      & 55.92 $\pm$ 0.59  & 57.34 $\pm$ 1.15 \\
900078              & 55.54 $\pm$ 1.89      & 53.31 $\pm$ 0.44  & 54.37 $\pm$ 1.78 \\
900085              & 48.54 $\pm$ 1.24      & 46.91 $\pm$ 1.03  & 49.87 $\pm$ 1.09 \\
900106              & 53.74 $\pm$ 1.05      & 52.16 $\pm$ 0.48  & 52.12 $\pm$ 1.39 \\
900152              & 50.87 $\pm$ 1.16      & 50.16 $\pm$ 0.76  & 51.35 $\pm$ 2.46 \\
900216              & 49.28 $\pm$ 1.37      & 47.52 $\pm$ 1.35  & 49.65 $\pm$ 3.33 \\
\hline
\end{tabular}
\end{table}

\begin{table}
\caption{Atmospheric parameters from \citet{Carretta2009a}}
\label{tab:Car} 
\centering                                    
\begin{tabular}{cccc}  
\hline\hline 
\multirow{3}{*}{ID} & Effective     & Surface       & Microturbulence \\ 
                    & temperature   & gravity       & velocity        \\ 
                    & [K]           & [cm~s$^{-2}$] & [km~s$^{-1}$]   \\ 
\hline
900073              & 4289          & 1.21          & 1.51 \\
900078              & 4294          & 1.23          & 1.65 \\
900085              & 4312          & 1.26          & 1.45 \\
900106              & 4370          & 1.37          & 1.72 \\
900152              & 4463          & 1.55          & 1.48 \\
900216              & 4585          & 1.76          & 1.51 \\
\hline
\end{tabular}
\end{table}

\subsection{Chemical Abundance Measurement} \label{sec:sub:abund}
Chemical abundance ratios were measured using the 2019NOV version of the local thermodynamic equilibrium (LTE) code MOOG \citep{Sneden1973}. 
Model atmospheres for each star were generated based on spherical MARCS models \citep{Gustafsson2008}, adopting atmospheric parameters, such as effective temperature, surface gravity, and microturbulence, as determined by \citet{Carretta2009a}, which are listed in Table~\ref{tab:Car}.
Abundances were measured through spectral synthesis of individual absorption lines using the {\em synth} driver of MOOG.

\begin{figure*}[t!]
\centering
   \includegraphics[width=0.95\textwidth]{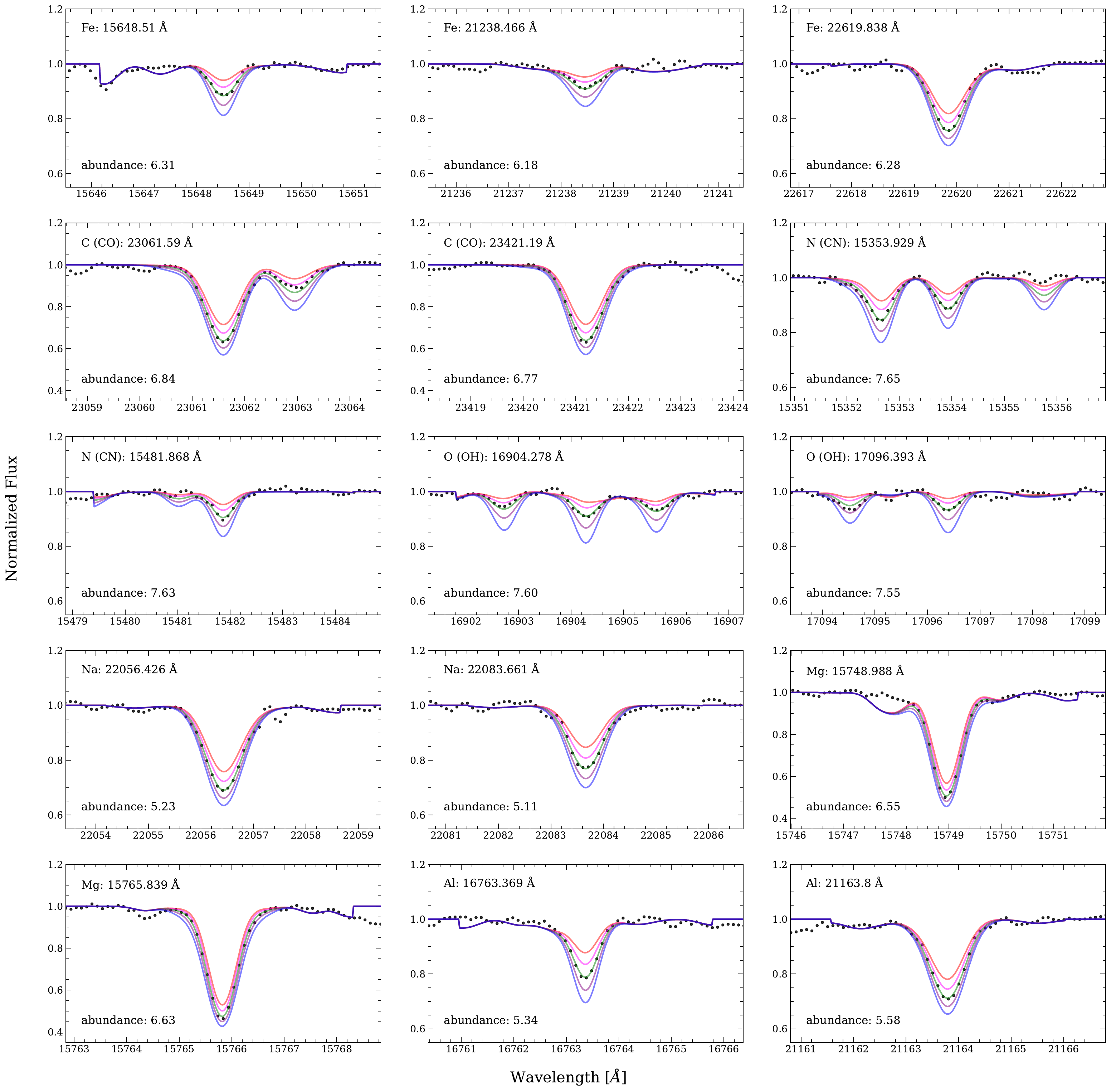}
     \caption{
     Example of spectral synthesis. 
     The observed spectrum of 900073 is displayed as black points, while the synthetic model spectra are represented by colored lines. 
     The green line indicates the best-fit model, while the red, magenta, purple and blue lines correspond to variations of $-0.4$, $-0.2$, $+0.2$, and $+0.4$~dex from the best-fit abundance, respectively.
     }
     \label{fig:synth}
\end{figure*}

Figure~\ref{fig:synth} illustrates examples of spectral synthesis for selected absorption lines, comparing the observed spectrum with synthetic models. 
Residuals between the observed and synthetic spectra in the central regions of each line were minimized to determine the best-fit abundances. 
The line list and associated parameters, including wavelength, excitation potential (EP), and oscillator strength ($\log{gf}$), were adopted from \citet{Lim2022} and \citet{Lim2024a}, which have been modified from \citet{Afsar2018}. 
The complete line list used in this study is provided in Table~\ref{tab:line}.

\begin{table}[t!]
\small
\setlength{\tabcolsep}{1pt}
\caption{Line list and information}
\label{tab:line} 
\centering                                                            
\begin{tabular}{@{\extracolsep{1pt}} l c c r @{\hskip 15pt} l c c r}   
\hline\hline 
$\lambda$ 	& Species	& EP		& $\log{gf}$	& $\lambda$	& Species	& EP		& $\log{gf}$  \\
\hline
$[\AA]$		&		& [eV]	& 			& $[\AA]$		&		& [eV]	& \\  
\hline
22056.426 & \ion{Na}{i} & 3.19 & 0.29    &	19917.191 & \ion{Ca}{i} & 1.90 & $-$2.48 \\
22083.661 & \ion{Na}{i} & 3.19 & $-$0.01 &	19933.727 & \ion{Ca}{i} & 3.91 & 0.16 \\
23348.423 & \ion{Na}{i} & 3.75 & 0.28    &	22607.944 & \ion{Ca}{i} & 4.68 & 0.43 \\
15024.997 & \ion{Mg}{i} & 5.11 & 0.36 	 &  22624.962 & \ion{Ca}{i} & 4.68 & 0.62 \\
15040.246 & \ion{Mg}{i} & 5.10 & 0.14    &	22626.723 & \ion{Ca}{i} & 4.68 & $-$0.32 \\
15047.714 & \ion{Mg}{i} & 5.11 & $-$0.34 &	15543.761 & \ion{Ti}{i} & 1.88 & $-$1.26 \\
15740.705 & \ion{Mg}{i} & 5.93 & $-$0.21 &	15602.842 & \ion{Ti}{i} & 2.27 & $-$1.59 \\
15748.988 & \ion{Mg}{i} & 5.93 & 0.14    &	21782.944 & \ion{Ti}{i} & 1.75 & $-$1.14 \\
15765.839 & \ion{Mg}{i} & 5.93 & 0.41    &	21897.376 & \ion{Ti}{i} & 1.74 & $-$1.44 \\
17108.631 & \ion{Mg}{i} & 5.39 & 0.06    &	22004.500 & \ion{Ti}{i} & 1.73 & $-$1.85 \\
21225.620 & \ion{Mg}{i} & 6.73 & $-$1.38 &	22211.238 & \ion{Ti}{i} & 1.73 & $-$1.75 \\
21458.865 & \ion{Mg}{i} & 6.52 & $-$1.32 &	22232.858 & \ion{Ti}{i} & 1.74 & $-$1.62 \\
22808.027 & \ion{Mg}{i} & 6.72 & $-$0.14 &	15680.060 & \ion{Cr}{i} & 4.70 & 0.15 \\
16718.957 & \ion{Al}{i} & 4.08 & 0.15    &	15860.210 & \ion{Cr}{i} & 4.70 & 0.00 \\
16763.369 & \ion{Al}{i} & 4.09 & $-$0.48 &	17708.730 & \ion{Cr}{i} & 4.39 & $-$0.51 \\
17699.050 & \ion{Al}{i} & 4.67 & $-$1.21 &	15194.490 & \ion{Fe}{i} & 2.22 & $-$4.75 \\
21093.078 & \ion{Al}{i} & 4.09 & $-$0.40 &	15207.526 & \ion{Fe}{i} & 5.39 & 0.08 \\
21163.800 & \ion{Al}{i} & 4.09 & $-$0.09 &	15343.788 & \ion{Fe}{i} & 5.65 & $-$0.69 \\
15960.080 & \ion{Si}{i} & 5.98 & 0.20    &	15648.510 & \ion{Fe}{i} & 5.43 & $-$0.70 \\
16060.021 & \ion{Si}{i} & 5.95 & $-$0.50 &	15662.013 & \ion{Fe}{i} & 5.83 & 0.07 \\
16094.797 & \ion{Si}{i} & 5.96 & $-$0.09 &	15761.313 & \ion{Fe}{i} & 6.25 & $-$0.16 \\
16163.714 & \ion{Si}{i} & 5.95 & $-$0.95 &	15858.657 & \ion{Fe}{i} & 5.58 & $-$1.25 \\
16215.691 & \ion{Si}{i} & 5.95 & $-$0.58 &	15980.725 & \ion{Fe}{i} & 6.26 & 0.72 \\
16241.851 & \ion{Si}{i} & 5.96 & $-$0.85 &	16009.610 & \ion{Fe}{i} & 5.43 & $-$0.55 \\
16434.929 & \ion{Si}{i} & 5.96 & $-$1.49 &	16153.247 & \ion{Fe}{i} & 5.35 & $-$0.73 \\
16680.770 & \ion{Si}{i} & 5.98 & $-$0.09 &	16165.029 & \ion{Fe}{i} & 6.32 & 0.75 \\
19928.919 & \ion{Si}{i} & 6.10 & $-$0.33 &	17420.825 & \ion{Fe}{i} & 3.88 & $-$3.52 \\
20343.887 & \ion{Si}{i} & 6.13 & $-$1.13 &	21178.155 & \ion{Fe}{i} & 3.02 & $-$4.24 \\
22537.686 & \ion{Si}{i} & 6.62 & $-$0.30 &	21238.466 & \ion{Fe}{i} & 4.96 & $-$1.37 \\
22665.777 & \ion{Si}{i} & 6.62 & $-$0.47 &	22257.107 & \ion{Fe}{i} & 5.06 & $-$0.82 \\
15163.090 & \ion{K}{i} & 2.67 & 0.55     &  22260.179 & \ion{Fe}{i} & 5.09 & $-$0.98 \\
15168.404 & \ion{K}{i} & 2.67 & 0.37     &  22392.878 & \ion{Fe}{i} & 5.10 & $-$1.32 \\
16136.823 & \ion{Ca}{i} & 4.53 & $-$0.67 &	22473.263 & \ion{Fe}{i} & 6.12 & 0.32 \\
16150.762 & \ion{Ca}{i} & 4.53 & $-$0.28 &	22619.838 & \ion{Fe}{i} & 4.99 & $-$0.51 \\
16155.236 & \ion{Ca}{i} & 4.53 & $-$0.77 &	23308.477 & \ion{Fe}{i} & 4.08 & $-$2.73 \\
16157.364 & \ion{Ca}{i} & 4.55 & $-$0.24 &	16310.501 & \ion{Ni}{i} & 5.28 & $-$0.02 \\
19815.017 & \ion{Ca}{i} & 4.62 & 0.40    &	16867.283 & \ion{Ni}{i} & 5.47 & $-$0.01 \\
\hline
\end{tabular}
\end{table}

\begin{table*}
\small
\setlength{\tabcolsep}{4pt}
\caption{Chemical abundance ratios and their measurement errors}
\label{tab:abund} 
\centering                                    
\begin{tabular}{ccccccc}
\hline\hline 
Abundance & \multirow{2}{*}{900073} & \multirow{2}{*}{900078} & \multirow{2}{*}{900085} & \multirow{2}{*}{900106} & \multirow{2}{*}{900152} & \multirow{2}{*}{900216} \\
ratio     &  &  &  &  &  &  \\
\hline
{[Fe/H]} & $-$1.27 $\pm$ 0.02 (17) & $-$1.25 $\pm$ 0.04 (12) & $-$1.28 $\pm$ 0.03 (18) & $-$1.29 $\pm$ 0.02 (15) & $-$1.29 $\pm$ 0.03 (15) & $-$1.28 $\pm$ 0.02 (14) \\
{[C/Fe]} & $-$0.37 $\pm$ 0.03 (15) & $-$0.98 $\pm$ 0.04 (13) & $-$0.47 $\pm$ 0.03 (15) & $-$0.52 $\pm$ 0.03 (16) & $-$0.62 $\pm$ 0.03 (14) & $-$0.38 $\pm$ 0.03 (15) \\
{[N/Fe]} &  1.04 $\pm$ 0.02 (14) &  1.27 $\pm$ 0.04 (13) &  0.57 $\pm$ 0.06 (9) &  0.63 $\pm$ 0.06 (6) &  0.94 $\pm$ 0.06 (12) &  0.83 $\pm$ 0.06 (7) \\
{[O/Fe]} &  0.16 $\pm$ 0.02 (11) & $-$0.13 $\pm$ 0.04 (8) &  0.30 $\pm$ 0.03 (12) &  0.30 $\pm$ 0.03 (11) &  0.12 $\pm$ 0.05 (10) &  0.43 $\pm$ 0.02 (8) \\
{[Na/Fe]} &  0.17 $\pm$ 0.04 (3) &  0.27 $\pm$ 0.04 (2) & $-$0.28 $\pm$ 0.06 (3) & $-$0.21 $\pm$ 0.05 (2) &  0.01 $\pm$ 0.06 (3) & $-$0.21 $\pm$ 0.05 (3) \\
{[Mg/Fe]} &  0.33 $\pm$ 0.05 (9) &  0.14 $\pm$ 0.06 (8) &  0.30 $\pm$ 0.04 (8) &  0.24 $\pm$ 0.03 (8) &  0.26 $\pm$ 0.06 (9) &  0.24 $\pm$ 0.03 (9) \\
{[Al/Fe]} &  0.30 $\pm$ 0.06 (4) &  0.68 $\pm$ 0.07 (4) & $-$0.02 $\pm$ 0.05 (5) & $-$0.02 $\pm$ 0.06 (4) &  0.37 $\pm$ 0.05 (5) & $-$0.01 $\pm$ 0.08 (4) \\
{[Si/Fe]} &  0.31 $\pm$ 0.03 (12) &  0.26 $\pm$ 0.05 (10) &  0.36 $\pm$ 0.04 (11) &  0.31 $\pm$ 0.03 (10) &  0.33 $\pm$ 0.03 (11) &  0.30 $\pm$ 0.03 (10) \\
{[K/Fe]} &  0.04 $\pm$ 0.06 (2) &  0.01 $\pm$ 0.04 (1) &  0.46 $\pm$ 0.03 (1) &  0.34 $\pm$ 0.07 (2) &  0.29 $\pm$ 0.06 (2) &  0.25 $\pm$ 0.03 (2) \\
{[Ca/Fe]} &  0.31 $\pm$ 0.06 (10) &  0.31 $\pm$ 0.06 (8) &  0.43 $\pm$ 0.05 (9) &  0.41 $\pm$ 0.04 (9) &  0.33 $\pm$ 0.05 (9) &  0.42 $\pm$ 0.04 (9) \\
{[Ti/Fe]} &  0.07 $\pm$ 0.02 (6) & $-$0.02 $\pm$ 0.05 (7) &  0.14 $\pm$ 0.04 (6) &  0.08 $\pm$ 0.03 (7) &  0.07 $\pm$ 0.04 (7) &  0.21 $\pm$ 0.03 (4) \\
{[Cr/Fe]} &  0.00 $\pm$ 0.02 (1) &  0.05 $\pm$ 0.04 (1) & $-$0.04 $\pm$ 0.09 (2) &  0.14 $\pm$ 0.02 (1) &  0.23 $\pm$ 0.03 (2) &  0.34 $\pm$ 0.03 (2) \\
{[Ni/Fe]} &  0.02 $\pm$ 0.15 (3) & $-$0.26 $\pm$ 0.08 (2) & $-$0.28 $\pm$ 0.04 (2) & $-$0.26 $\pm$ 0.04 (2) & $-$0.25 $\pm$ 0.08 (2) & $-$0.03 $\pm$ 0.03 (2) \\
\hline
\addlinespace[2mm]
\end{tabular}
\parbox{\textwidth}{\small \textbf{Note.} Numbers in parentheses indicate the number of spectral lines used for abundance measurements.}
\end{table*}

As described in Section~\ref{sec:intro}, this study focused on key elements relevant to investigating multiple stellar populations in GCs, including Na, O, Mg, and Al, as well as Fe. 
The [Fe/H] abundance ratio was initially measured using 20 \ion{Fe}{i} absorption lines across the spectral range. 
However, due to low SNR and poor fits for certain lines, the [Fe/H] ratio was ultimately derived from approximately 15 lines per star. 
Final [Fe/H] values were calculated as the mean of these measurements, with measurement uncertainties estimated as the standard error of the mean.

The abundances of C, N, and O were measured from CO, CN, and OH molecular bands, respectively. 
As the abundances of these elements are interconnected through molecular features, they were determined iteratively. 
First, O was measured from OH bands, followed by C from CO bands, and finally N from CN bands. 
During this process, the abundances of other elements in the model atmosphere were updated iteratively. 
The molecular bands used for this analysis included: CH bands near 22966.62, 22972.33, 22978.44, 23015.00, 23023.52, 23051.47, 23061.59, 23083.04, 23106.10, 23118.23, 23274.53, 23288.74, 23351.41, 23384.44, 23421.19, and 23434.27~$\AA$; CN bands near 15011.38, 15021.15, 15076.32, 15080.21, 15134.12, 15221.84, 15242.39, 15318.74, 15328.50, 15353.93, 15397.43, 15410.56, 15447.10, 15466.24, 15481.87, and 15495.26~$\AA$; and OH bands near 15755.52, 15756.53, 16052.77, 16055.47, 16192.13, 16251.66, 16443.83, 16526.25, 16904.28, 17096.39, 17100.45, and 17316.13~$\AA$. 
After three iterations, fixed abundance ratios for each element were determined, ensuring they were independent of variations in the abundances of other elements.

Abundances of other elements, including Na, Mg, Al, Si, Ca, and Ti, were measured through spectral synthesis of individual absorption lines listed in Table~\ref{tab:line}.
Table~\ref{tab:abund} summarizes the measured [Fe/H] and other abundance ratios ([X/Fe]) for each target, based on the solar abundance scale of \citet{Asplund2009}. 
Measurement uncertainties, calculated as the standard error of the mean from multiple lines, are also included.

\section{Results of Chemical Abundance Measurements} \label{sec:result}
\subsection{Comparison with Optical Spectroscopy} \label{sec:sub:comp}
The detailed chemical properties of multiple stellar populations in GCs have traditionally been studied using optical spectroscopy. 
However, high-resolution NIR spectroscopy has recently been employed in several studies to investigate GCs, particularly in the inner Galaxy \citep[e.g.,][]{Taylor2022, Lim2024b}. 
To evaluate the effectiveness of NIR spectroscopy in GC research, we compared our chemical abundance measurements with the optical spectroscopy results reported by \citet{Carretta2009a}.

\begin{figure}[t!]
\centering
   \includegraphics[width=0.48\textwidth]{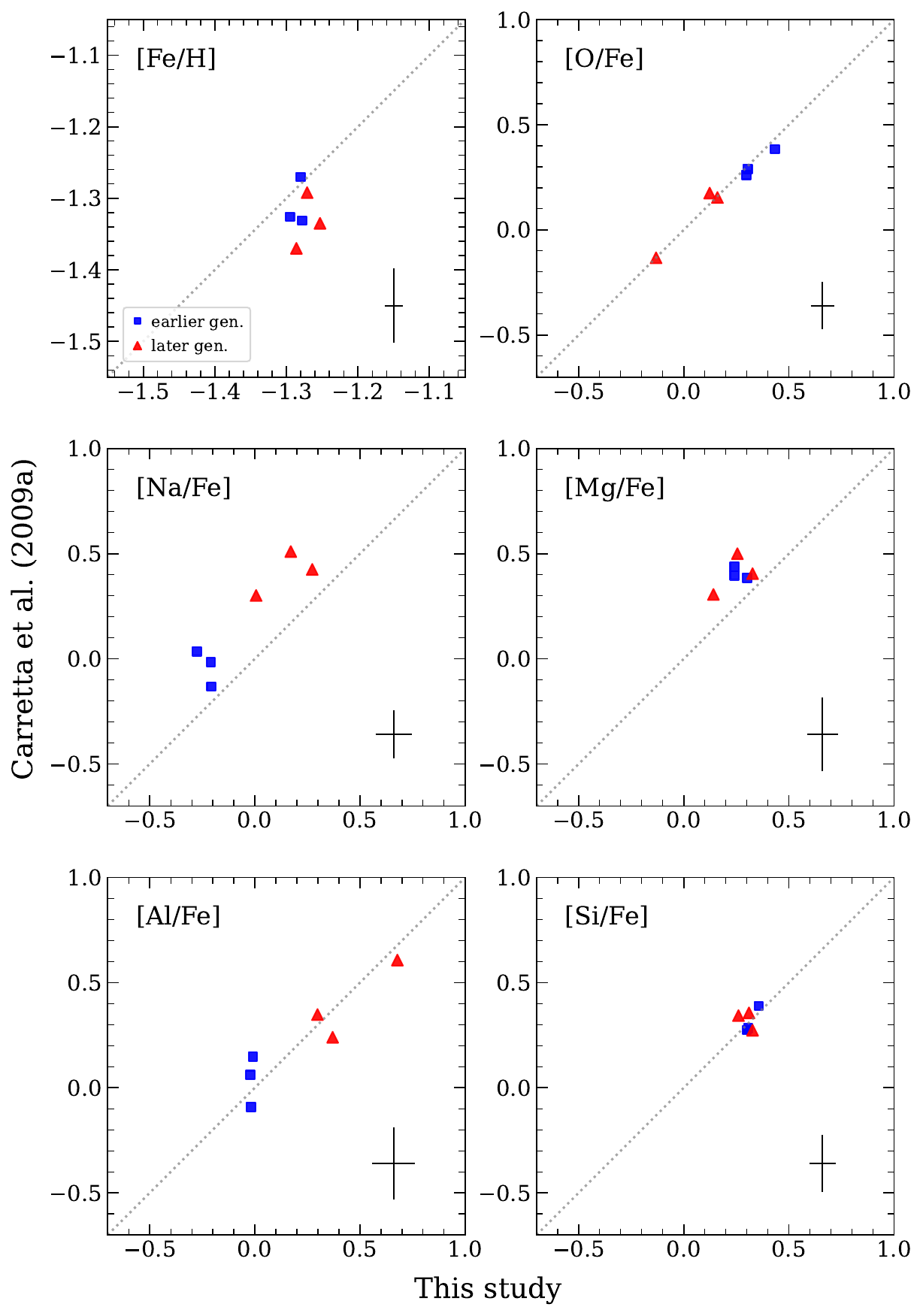}
     \caption{
     Comparison of chemical abundance ratios between NIR measurements (this study) and optical measurements from \citet{Carretta2009a}. 
     Blue squares represent earlier-generation stars, while red triangles indicate later-generation stars. 
     The average measurement error for each element from both studies is displayed in the lower right corner of each panel. 
     Although the measurements generally show good agreement, systematic differences are observed in the [Fe/H], [Na/Fe], and [Mg/Fe] abundance ratios.
     }
     \label{fig:comp}
\end{figure}

Figure~\ref{fig:comp} presents a direct comparison of the chemical abundance ratios for Fe, O, Na, Mg, Al, and Si, which are common to both studies. 
Overall, the abundance ratios derived from NIR and optical spectroscopy are consistent, with minor offsets. 
The differences in [Fe/H] between the two studies are less than 0.09~dex, with our estimates being, on average, 0.04~dex higher than those reported by \citet{Carretta2009a}. 
Similarly, the [O/Fe], [Al/Fe], and [Si/Fe] abundance ratios show good agreement, with mean differences of 0.03~dex, 0.09~dex, and 0.04~dex, respectively, and no significant systematic offsets.

However, larger offsets are observed for [Na/Fe] and [Mg/Fe], where our measurements are generally lower than those reported by \citet{Carretta2009a}. 
The typical differences are 0.23~dex for [Na/Fe] and 0.15~dex for [Mg/Fe]. 
These discrepancies likely reflect systematic differences between optical and NIR spectroscopic analyses, potentially due to the selection of absorption lines.
Notably, our previous NIR spectroscopic observations also showed underestimations of the Mg abundance compared to previously reported values \citep[see][]{Lim2024b}.

Despite these systematic differences, the strong correlation between measurements from the optical and NIR bands underscores the reliability of NIR spectroscopy for GC studies. 
These findings highlight the potential of NIR spectroscopy as a reliable counterpart for chemical abundance analysis in GCs.

\subsection{Application to Multiple Stellar Population Studies} \label{sec:sub:na_o}
When studying multiple stellar populations in GCs, the relative differences in abundance ratios between populations are more significant than the absolute abundance ratios of individual elements. 
A hallmark of multiple stellar populations is the presence of anti-correlations among specific elements, as discussed in Section~\ref{sec:intro}. 
Later-generation stars in GCs are typically enriched in N, Na, and Al, while being depleted in C, O, and Mg, relative to earlier-generation stars. 
Based on these chemical characteristics, GC stars can be classified into primordial (earlier) and enriched (later) populations.

\begin{figure}[t!]
\centering
   \includegraphics[width=0.4\textwidth]{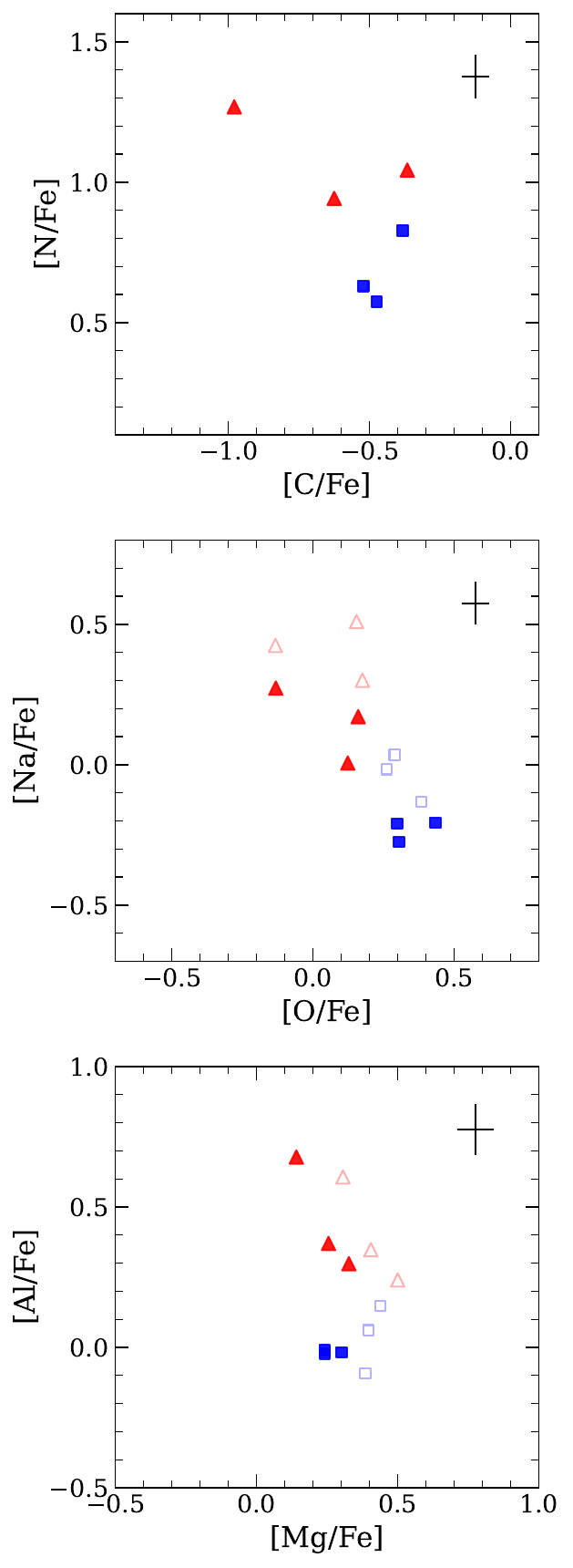}
     \caption{
     Chemical patterns of multiple stellar populations in the N-C (upper), Na-O (middle), and Mg-Al (lower) planes. 
     The symbols are the same as in Figure~\ref{fig:comp}, with open symbols representing abundance ratios from \citet{Carretta2009a}. 
     Open symbols are not included in the top panel because \citet{Carretta2009a} did not measure C and N abundance ratios.
     The six sample stars clearly show N-C and Na-O anti-correlations, while no significant difference is observed in the [Mg/Fe] abundance ratio. 
     In addition, the early and later stellar populations are distinctly separated in these plots.
     }
     \label{fig:Na_O}
\end{figure}

Our sample includes three primordial and three enriched population stars, as identified by \citet{Carretta2009a}. 
Figure~\ref{fig:Na_O} shows the distribution of these stars on the N-C, Na-O, and Mg-Al planes, illustrating the distinction between the two populations.
Consistent with the agreement between our NIR-based abundances and those of \citet{Carretta2009a}, the two populations are clearly separated in these diagrams.
Later-generation stars (900073, 900078, and 900152) exhibit higher [N/Fe], [Na/Fe], and [Al/Fe] abundance ratios, as well as lower [C/Fe] and [O/Fe] ratios, compared to earlier-generation stars (900085, 900106, and 900216).

For Mg abundances, no clear separation is observed between the two populations, consistent with the findings of \citet{Carretta2009a} (see open symbols in Figure~\ref{fig:Na_O}). 
When abundance ratios from \citet{Carretta2009a} are overplotted on the Na-O and Mg-Al planes, both their study and ours successfully distinguish the two populations, despite the discrepancies in Na and Mg abundances discussed in Section~\ref{sec:sub:comp}. 
These results confirm that high-resolution NIR spectroscopy can reliably identify multiple stellar populations in GCs, providing confidence levels comparable to those achieved through high-resolution optical spectroscopy.

In addition to reproducing previous measurements, we measured the abundances of C and N, which were not reported by \citet{Carretta2009a}. 
While these elements can be determined from optical CN and CH bands at 3883~$\AA$ and 4300~$\AA$, respectively \citep[e.g.,][]{Lardo2012, Kim2022}, high-resolution NIR spectroscopy provides numerous spectral features that allow for more precise measurements. 
\citet{Hong2021}, who studied the peculiar GC NGC~2808, emphasized the importance of the N-C (or CN-CH) anti-correlation in GC studies. 
They found that [N/Fe] is particularly effective for subdividing earlier-generation stars, while [Na/Fe] is more reliable for distinguishing later-generation stars due to differences in chemical yields. 
Thus, combining Na-O and N-C abundance patterns is highly advantageous for identifying multiple stellar populations.

In the upper panel of Figure~\ref{fig:Na_O}, two stars—one from the earlier population and one from the later population—appear somewhat enhanced in both [C/Fe] and [N/Fe] relative to the other four stars. 
In addition, one later-generation star is distinctly separated, showing a higher [N/Fe] ratio. 
However, due to the small sample size, it is challenging to draw definitive conclusions from this distribution alone.

\begin{figure}[t!]
\centering
   \includegraphics[width=0.4\textwidth]{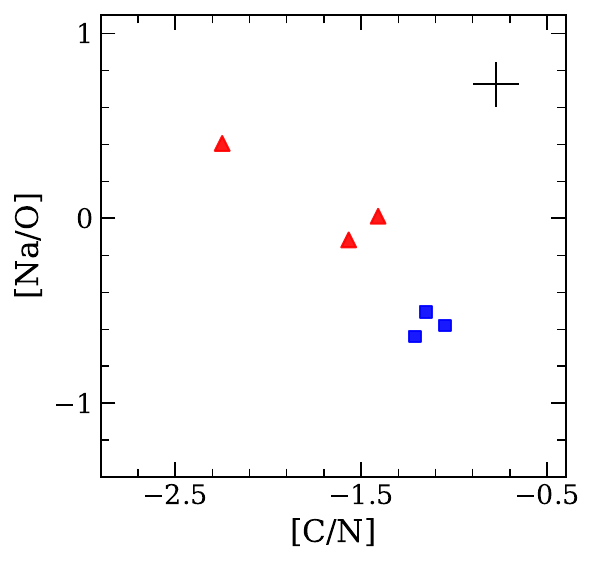}
     \caption{
     Chemical abundance distribution of sample stars in the [Na/O] versus [C/N] plane. 
     The symbols are consistent with those in Figure~\ref{fig:comp}. 
     Later-generation stars can be further subdivided into two groups with lower and intermediate [C/N] abundance ratios, while earlier-generation stars are tightly clustered. 
     This plot highlights the effectiveness of NIR spectroscopy in studying multiple stellar populations in GCs.
     }
     \label{fig:NaO_CN}
\end{figure}

Figure~\ref{fig:NaO_CN} shows the distribution of [Na/O] versus [C/N] abundance ratios for the six sample stars. 
Notably, this plot reveals more distinct sub-grouping compared to the abundance planes shown in Figure~\ref{fig:Na_O}. 
The three earlier-generation stars are tightly clustered, while the later-generation stars are divided into two groups with low and intermediate [C/N] abundance ratios. 
One later-generation star, which exhibits a higher [Na/O] ratio and a lower [C/N] ratio, likely belongs to the extreme stellar population observed in some GCs, such as NGC~2808 \citep{Carlos2023}. 
Evidence for this extreme population is also hinted at by the depletion of O and the enhancement of Al in this star (see Figure~\ref{fig:Na_O}).

This study does not aim to investigate the origins of multiple stellar populations in M5, as the primary objective is to evaluate the utility of high-resolution NIR spectroscopy for GC studies. 
Nonetheless, our results demonstrate that high-resolution NIR spectroscopy is a powerful tool for investigating multiple stellar populations in GCs, providing results that are both comparable to and complementary with those of high-resolution optical spectroscopy.

\section{Discussion} \label{sec:discussion}
We have demonstrated that high-resolution NIR spectroscopic data, obtained with the IGRINS-2 spectrograph on the Gemini-North telescope, can produce results comparable to those of high-resolution optical spectroscopy for studies of multiple stellar populations in GCs. 
The measured abundance ratios of six sample stars show good agreement with those reported by \citet{Carretta2009a}, with minor offsets observed in the Na and Mg abundances. 
Furthermore, our results effectively distinguish multiple stellar populations in the N-C, Na-O, and Mg-Al planes.
The derived chemical properties of M5 are also consistent with previous studies.
The large star-to-star variations in the abundances of Na, O, and Al, together with the small scatter in other elements, including Fe, align with the findings of \citet{Ivans2001} and \citet{Koch2010}.
In addition, the distributions and ranges of [C/Fe] and [N/Fe] ratios are comparable to those reported by \citet{Smith1997}, further validating the reliability of our results.

Looking ahead, we anticipate that NIR spectroscopic data obtained from IGRINS-2 and other advanced NIR spectrographs will play an increasingly important role in the study of stellar populations in Milky Way GCs and stellar clusters in other galaxies. 
This is especially relevant as the applications of NIR spectroscopy broaden through increasingly diverse and detailed research efforts.
Below, we highlight the key advantages of high-resolution NIR spectroscopy and address potential challenges for future studies.

\subsection{Advantages of NIR Spectroscopy} \label{sec:sub:adv}
One of the most significant advantages of NIR spectroscopy is its resilience to interstellar reddening, making it particularly valuable for uncovering and studying newly discovered GCs in the inner Galaxy, where extinction is severe \citep[see, e.g.,][]{Bica2024}.
For example, the E(B-V) reddening value for the bulge GC Terzan~5 is 2.28 mag, compared to only 0.03 mag for M5, a halo GC \citep{Harris2010}. 
This large difference implies that stars in Terzan~5 appear approximately six magnitudes fainter in the V-band than stars with identical luminosities in M5, whereas the corresponding difference in the K-band is only about 0.7 magnitudes. 
Such significant extinction in the optical bands necessitates much longer observation times, making NIR spectroscopy essential for chemical studies of GCs in the Galactic bulge.

Another notable advantage of high-resolution NIR spectroscopy is its capacity to measure the chemical abundances of C, N, and O elements, which are critical for understanding multiple stellar populations in GCs. 
In the optical wavelength range, O abundance measurements are limited to a few spectral features, such as the forbidden [\ion{O}{i}] lines at 6300.3 and 6363.8~$\AA$.
These lines are prone to telluric contamination and blending with nearby \ion{Ni}{i} lines, requiring careful analysis. 
Similarly, the strong CN and CH molecular bands for N and C abundances are located at the blue edge of optical spectra, where observing efficiency is highly reduced.
In contrast, the NIR wavelength range includes numerous CH, CN, and OH molecular features, allowing for more reliable and precise measurements of C, N, and O abundances. 
As shown in Section~\ref{sec:sub:abund}, we measured these elements using more than 10 spectral features per element, with measurement uncertainties comparable to or smaller than those of other elements (see Table~\ref{tab:abund}). 
This capability not only strengthens the utility of NIR spectroscopy for studying multiple stellar populations in GCs but also broadens its applications to investigating the stellar chemical abundances of volatile elements, which are essential for understanding planet-hosting stars \citep[e.g.,][]{Yun2024}.

\begin{figure}[t!]
\centering
   \includegraphics[width=0.37\textwidth]{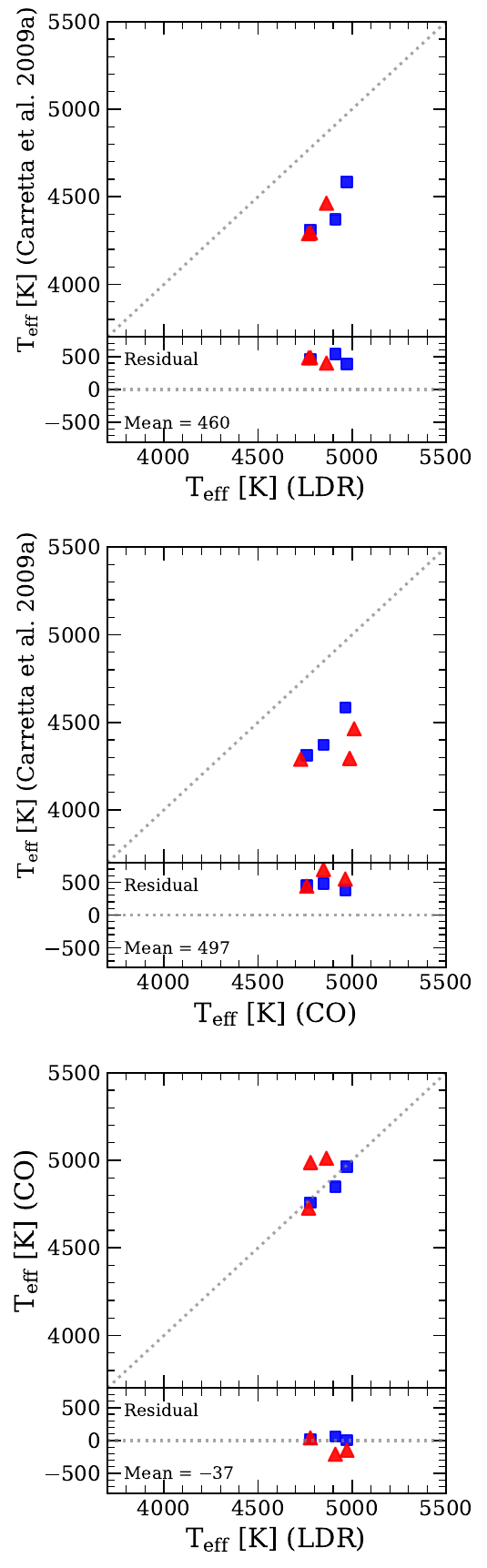}
     \caption{
     Comparison of ${T_{\rm eff}}$ values derived from different methods. 
     The ${\rm T_{eff}}$ estimates derived solely from NIR spectroscopic data using both LDRs and the CO-overtone band are higher than those reported by \citet{Carretta2009a} (upper and middle panels). 
     However, the two NIR-based methods show good agreement with each other (bottom panel), with a standard deviation of 110~K.
     }
     \label{fig:teff}
\end{figure}

\subsection{Future Challenges of NIR Spectroscopy} \label{sec:sub:chall}
A critical challenge in chemical abundance measurements is the accurate determination of atmospheric parameters for model atmospheres. 
These parameters, particularly effective temperature (${T_{\rm eff}}$), have a significant influence on the measurement of abundance ratios.
While both photometric and spectroscopic methods can be used to derive these parameters, traditional spectroscopic approaches are not applicable in the NIR region due to the absence of \ion{Fe}{ii} lines. 
Photometric methods, meanwhile, are often unreliable for stars in the inner Galaxy because of severe and variable reddening effects \citep[see][for more details]{Lim2022}. 
Alternative approaches, such as the use of line-depth ratios \citep[LDRs;][]{Afsar2023} and CO-overtone bands \citep{Park2018}, have been proposed but require further validation to establish their reliability.

In this study, we adopted the atmospheric parameters, ${\rm T_{eff}}$, surface gravity ($\log{g}$), and microturbulence velocity ($\xi_{\rm t}$), from \citet{Carretta2009a}, as M5 is located in the Galactic halo with minimal reddening, allowing relatively accurate parameter determination through photometric methods. 
However, when we derived ${T_{\rm eff}}$ values solely from NIR spectroscopic data, discrepancies emerged. 
As shown in Figure~\ref{fig:teff}, ${T_{\rm eff}}$ values derived using the LDR method \citep{Afsar2023} and the CO-band relation \citep{Park2018} were, on average, 460~K and 497~K higher, respectively, than those reported by \citet{Carretta2009a}. 
Both methods are valid for dwarf and giant stars with ${T_{\rm eff}}$ ranges of 3200 $-$ 5500~K and 3000 $-$ 6000~K, placing our sample stars within their applicable ranges. 
Notably, the two NIR-based methods were in good agreement with each other, exhibiting a standard deviation of only 110~K (see the bottom panel of Figure~\ref{fig:teff}).

The exact cause of these discrepancies remains unclear but may be attributed to differences in luminosity classes or the blending of spectral features \citep[see Section~5 in][]{Afsar2023}. 
Nevertheless, a promising strategy to tackle this problem is the grid-fitting technique utilizing synthetic spectra (Yun et al., in preparation), and we carried out a simple comparison exercise. To briefly summarize, we initially created a grid of model atmospheres from Kurucz's NEWODF models \citep{Castelli2004} and solar relative abundances from \citet{Asplund2005} using the ATLAS9 model atmosphere codes from \citet{Castelli2004}. These model atmospheres assume one-dimensional LTE with plane-parallel line-blanketed model structure and an increase in alpha-element abundances by $+$0.4 dex for stars with [Fe/H] $\leq$ --1.0 and $+$0.3, $+$0.2, $+$0.1 dex for [Fe/H] = --0.75, --0.5, --0.25, respectively. The grid ranges from 3500 K $\leq$ $T_{\rm efff}$ $\leq$ 8,000 K in 250 K increments, 0.0 $\leq$ $\log{g}$ $\leq$ 5.0 in 0.25 dex increments, and --4.0 $\leq$ [Fe/H] $\leq$ +1.0 in 0.25 dex increments.

\begin{figure}[t!]
\centering
   \includegraphics[width=0.45\textwidth]{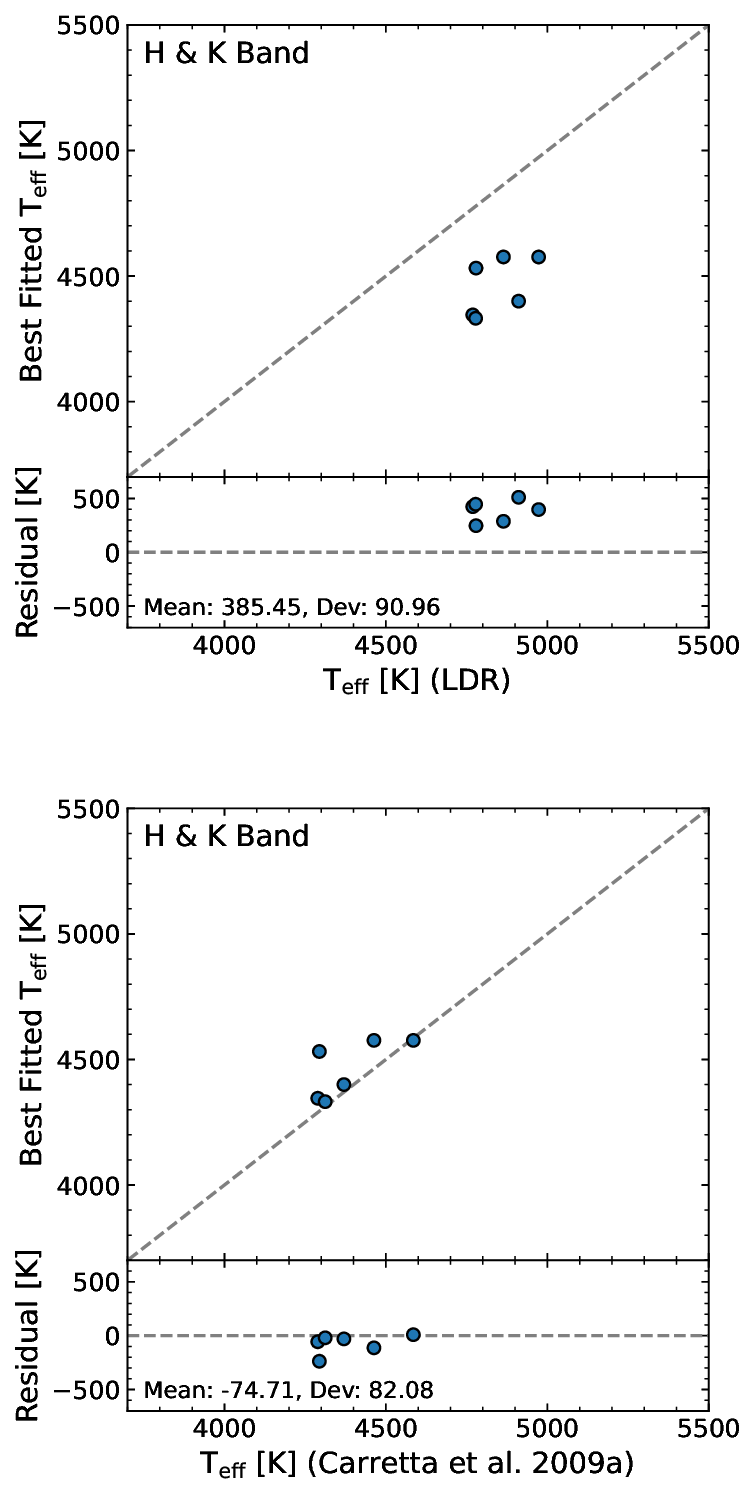}
     \caption{
     Comparison of ${T_{\rm eff}}$ between the LDR (top) and \citet{Carretta2009a} (bottom) and the grid-fitting method (bottom). The template matching approach shows a better agreement with a mean offset of --74 K with a standard deviation of 82 K.
     }
     \label{fig:teff_grid}
\end{figure}

We subsequently employed $synthe$ codes to generate synthetic spectra after incorporating the line lists from R. Kurucz website (http://kurucz.harvard.edu/). The wavelength range covers 14,000 to 25,000 \AA, encompassing the H- and K-bands. We adopted a straightforward formula for determining $\xi_{\rm t}$ for each spectrum, as per \citep{Lee2013}: $\xi_{\rm t}$ [km s$^{-1}$] = --0.345 $\cdot$ $\log{g}$ = $+$ 2.225. After inputting IGRINS-2 M5 spectra and synthetic spectra at identical resolution and normalizing both spectra, we explored the grid of synthetic spectra to identify the best-fitting model parameters by reducing the disparity between the normalized observed and synthetic spectra. Figure~\ref{fig:teff_grid} illustrates the outcomes of the grid-fitting process. The upper panel compares with the LDR method, while the lower panel compares with \citet{Carretta2009a}. We observe a good agreement with \citet{Carretta2009a}, showing a mean difference of --74 K and a standard deviation of 82 K. In contrast, the empirical method shows a mean offset of 385 K, a similar scale within the error as in Figure~\ref{fig:teff}. This exercise clearly suggests that the template matching method reproduces the temperature closer to the optical estimate than the empirical approaches, and a more thorough study on this should be carried out in the near future.


Another major challenge is the need for an expanded spectral line list for chemical abundance measurements in the NIR region. 
While our study and others have successfully measured abundances for various elements, the number of useful lines in the NIR remains significantly smaller than in the optical. 
Increasing the number of available lines would enhance the statistical precision of abundance measurements. 
Moreover, identifying line lists for new elements, such as neutron-capture elements, is crucial, as these are key to understanding the formation and evolution of GCs and individual stars \citep[e.g.,][]{Yong2014, Schiappacasse-Ulloa2023, Nandakumar2024}.

Furthermore, accurate information on EP and $\log{gf}$ values for absorption lines is essential to reduce line-to-line abundance variations and systematic discrepancies between NIR and optical spectroscopy. 
Non-LTE corrections for NIR spectral lines also need to be considered. 
For example, \citet{Lim2022} suggested that non-LTE effects might explain discrepancies in abundances derived from the H- and K-bands. 
Although some studies have explored non-LTE effects in the NIR \citep[e.g.,][]{Bergemann2012}, more comprehensive analyses are required. 
Improved line information and non-LTE corrections will significantly enhance the precision and broader applicability of NIR spectroscopy for stellar studies.

Finally, while IGRINS and IGRINS-2 spectrographs have primarily been utilized for exoplanet atmosphere studies, they hold substantial potential as leading instruments for high-resolution NIR stellar spectroscopy. 
As the challenges discussed above are addressed and resolved, these instruments will become even more powerful tools for advancing our understanding of GCs and stellar populations.



\acknowledgments
We thank the reviewer for a number of helpful suggestions.
D.L. and Y.W.L. acknowledge support from the National Research Foundation (NRF) of Korea to the Center for Galaxy Evolution Research (RS-2022-NR070872 and RS-2022-NR070525). 
Y.S.L. acknowledges support from the NRF of Korea grant funded by the Ministry of Science and ICT (RS-2024-00333766).
S.H.C. acknowledges support from the NRF of Korea grant funded by the Korea government (MSIT; NRF-2021R1C1C2003511).
This research was partly supported by the Korea Astronomy and Space Science Institute (KASI) under the R\&D program (Project No. 2025-1-860-02, Project No. 2025-1-868-02) supervised by the Korea AeroSpace Administration.
This work used the Immersion Grating Infrared Spectrograph 2 (IGRINS-2) developed and built by a collaboration between KASI and the International Gemini Observatory.
Based on observations obtained at the international Gemini Observatory, a program of NSF NOIRLab, which is managed by the Association of Universities for Research in Astronomy (AURA) under a cooperative agreement with the U.S. National Science Foundation on behalf of the Gemini Observatory partnership: the U.S. National Science Foundation (United States), National Research Council (Canada), Agencia Nacional de Investigaci\'{o}n y Desarrollo (Chile), Ministerio de Ciencia, Tecnolog\'{i}a e Innovaci\'{o}n (Argentina), Minist\'{e}rio da Ci\^{e}ncia, Tecnologia, Inova\c{c}\~{o}es e Comunica\c{c}\~{o}es (Brazil), and Korea Astronomy and Space Science Institute (Republic of Korea).
This work was enabled by observations made from the Gemini North telescope, located within the Maunakea Science Reserve and adjacent to the summit of Maunakea. We are grateful for the privilege of observing the Universe from a place that is unique in both its astronomical quality and its cultural significance.




\bibliography{export-bibtex}






\end{document}